\documentclass[11pt,letterpaper]{article}

\usepackage[T1]{fontenc}
\usepackage[utf8]{inputenc}
\usepackage[english]{babel}
\usepackage{lmodern}
\usepackage[margin=1in]{geometry}
\usepackage{microtype}
\usepackage{amsmath,amssymb,bm}
\usepackage{graphicx}
\usepackage{booktabs}
\usepackage[font=small,labelfont=bf]{caption}
\usepackage{float}
\usepackage[authoryear,round]{natbib}
\usepackage{xurl}
\usepackage[hidelinks]{hyperref}

\floatstyle{ruled}
\newfloat{algorithm}{tbp}{loa}
\floatname{algorithm}{Algorithm}
\floatstyle{plain}

\hypersetup{
  pdftitle={DeepSurrogate: An Interpretable Artificial Intelligence System for Efficient Modeling of Functional Surrogates for High-Fidelity Computer Models},
  pdfauthor={Yeseul Jeon, Rajarshi Guhaniyogi, Aaron Scheffler, Devin Francom, Donatella Pasqualini}
}
\allowdisplaybreaks[1]

\newcommand{\bb}{ {\boldsymbol b} }
\newcommand{\bB}{ {\boldsymbol B} }

\newcommand{\bD}{ {\boldsymbol D} }

\newcommand{\bF}{ {\boldsymbol F} }

\newcommand{\bI}{ {\boldsymbol I} }

\newcommand{\br}{ {\boldsymbol r} }
\newcommand{\bR}{ {\boldsymbol R} }
\newcommand{\bs}{ {\boldsymbol s} }

\newcommand{\bw}{ {\boldsymbol w} }
\newcommand{\bW}{ {\boldsymbol W} }
\newcommand{\bx}{ {\boldsymbol x} }

\newcommand{\by}{ {\boldsymbol y} }

\newcommand{\bz}{ {\boldsymbol z} }

\newcommand{\bbeta}   { {\boldsymbol \beta} }

\newcommand{\bet}     { {\boldsymbol \eta} }
\newcommand{\btheta}  { {\boldsymbol \theta} }
\newcommand{\bTheta}  { {\boldsymbol \Theta} }

\newcommand{\bmu}     { {\boldsymbol \mu} }

\newcommand{\bSigma}  { {\boldsymbol \Sigma} }

\newcommand{\bphi}    { {\boldsymbol \phi} }
\newcommand{\bPhi}    { {\boldsymbol \Phi} }

\newcommand{\bzero}  { {\boldsymbol 0} }

\title{DeepSurrogate: An Interpretable Artificial Intelligence System
for Efficient Modeling of Functional Surrogates for High-Fidelity Computer Models}
\author{%
  Yeseul Jeon\textsuperscript{1}\quad
  Rajarshi Guhaniyogi\textsuperscript{1}\quad
  Aaron Scheffler\textsuperscript{2}\\[0.4em]
  Devin Francom\textsuperscript{3}\quad
  Donatella Pasqualini\textsuperscript{3}\\[0.8em]
  \small\textsuperscript{1}Department of Statistics, Texas A\&M University\\
  \small\textsuperscript{2}Department of Epidemiology and Biostatistics,\\
  \small University of California, San Francisco\\
  \small\textsuperscript{3}Los Alamos National Laboratory
}
\date{June 29, 2026}

\begin{document}
\maketitle

\begin{abstract}
Developing statistical surrogates for high-fidelity, computationally intensive simulators is essential for complex problems in science and engineering. This article introduces \emph{DeepSurrogate}, an artificially intelligent (AI) surrogate model tailored for functional outputs with vector-valued inputs. The model represents the input-output relationship as an infinite collection of spatially indexed functions, each capturing how inputs influence the output at a given location. These functions are decomposed into two components: one modeling nonlinear dependence on input variables and the other capturing spatial correlations across output locations. Both components are modeled using deep neural networks (DNNs), allowing it to jointly model spatial structure and nonlinear input-output relationships. A key feature of DeepSurrogate is its ability to quantify predictive uncertainty via Monte Carlo dropout, while embedding deep neural networks within a statistical framework that enhances interpretability. To our knowledge, it is the first \emph{interpretable DNN} architecture for function-on-scalar regression quantifying uncertainty. The method offers full predictive inference with approximately 50,000 spatial locations and 20 simulation runs, with training and testing completed in under 10 minutes. Its effectiveness is demonstrated through synthetic simulations and a large-scale application involving the Sea Lake and Overland Surge from Hurricanes (SLOSH) simulations. A Python implementation is publicly available.
\end{abstract}

\section{Introduction}
Scientific analysis in environmental applications often relies on simulations from computer models generating high-resolution functional data, such as climate and weather systems \citep{petersen2019evaluation,borge2008comprehensive}. These simulators face key challenges in terms of significant computational resources and run times, as well as characterizing uncertainty in input attributes for reliable conclusions \citep{marrel2011global, el2002encyclopedia}. Input uncertainty can be addressed intrusively, by modifying the simulation model to propagate uncertainty \citep{xiu2002wiener, najm2009uncertainty}, or non-intrusively, by treating the simulator as a black box and running it at various settings \citep{santner2003design, higdon2004combining}. Non-intrusive methods are more practical as they avoid altering the simulator's structure. However, this
approach involves conducting numerous runs of the simulator with various input settings, and each of these runs comes with substantial computational costs for complex, high-fidelity (expensive) simulators. As a result, the process of simulating multiple runs is significantly expensive for many simulators and often infeasible in terms of computational resources. To mitigate this, statistical surrogates (or emulators) are often built to approximate the simulator. Once trained, these emulators enable faster predictions, facilitating studies of model response and parameter uncertainty.

This manuscript is motivated by physics-based simulations that use a hurricane's track and hurricane-related attributes as inputs to generate output of a storm surge function or flood depth measurements above normal levels (e.g., water depth above ground at land locations), over a large spatial grid. The primary goal is to model the maximum (in time) flood depth across the entire spatial domain as a functional output, enabling predictive inference of this function for untested hurricane tracks and attributes, quantifying uncertainty, and achieving these results significantly faster than actually running a large number of physics-based simulations for uncertainty quantification (UQ).

The scientific problem can be formulated within a function-on-scalar (FoS) regression framework, where much of the existing literature emphasizes functional linear models (FLM). In these models, functionally varying coefficients corresponding to scalar inputs are employed to capture and explain variations in a functional response. Early approaches to FoS regression primarily relied on basis function expansions, where both the response and coefficient functions are expressed using a finite set of basis functions \citep{reiss2007functional}. This transformation reduces the infinite-dimensional problem into a manageable finite-dimensional one, enabling the use of standard regression techniques to estimate the coefficients effectively. Another widely adopted method is penalized functional regression, which incorporates smoothness penalties (e.g., roughness penalties on derivatives of coefficient functions) during estimation \citep{goldsmith2011penalized}. Additionally, functional principal component regression (FPCR) has been employed to reduce the dimensionality of functional responses \citep{yao2006penalized, goldsmith2015generalized}. In FPCR, principal components of the functional responses are extracted, and regression is performed on these reduced components, preserving essential information while reducing computational complexity. Beyond the FLM framework, several nonlinear approaches have been developed to address more intricate regression scenarios. For example, \cite{zhang2015varying} combined the FoS regression framework with additive models, removing the linearity assumption between inputs and the functional response. This enhancement provides greater flexibility in modeling complex relationships. Expanding on this idea, \cite{scheipl2015functional} proposed the functional additive mixed (FAM) model, offering a more generalized extension of the additive approach.

Although these traditional methods provide uncertainty estimates for model parameters, they are not specifically designed to quantify uncertainty in the prediction of functional outcomes. Bootstrap-based approaches have been proposed for predictive uncertainty quantification \citep{kirk2009gaussian}, but they lack theoretical guarantees when applied to FoS regression for estimation of functions. Additionally, applying conformal prediction in this context is challenging, as it relies on the exchangeability of responses, an assumption that does not hold in standard FoS regression settings. To address this limitation, Bayesian approaches have gained traction in the FoS regression framework \citep{morris2015functional}. A class of Bayesian approaches employs varying coefficient models, with varying coefficients capturing the complex non-linear association between functional outcome and scalar input \citep{andros2024robust}. Although Gaussian Process (GP) priors are an attractive choice for varying coefficients due to their flexibility, smooth interpolation capabilities, and built-in uncertainty quantification for predictive distributions \citep{santner2003design, gramacy2020surrogates}, they suffer from significant computational limitations, primarily due to the need to invert large covariance matrices, which hampers their scalability for large datasets.

Developing surrogates with large functional data in computer experiments has garnered increasing interest, leading to a burgeoning literature on scalable methods, many of which draw from advancements in scalable spatial modeling. These approaches include multi-resolution methods \citep{guhaniyogi2020large}, local Gaussian processes \citep{gramacy2015local}, deep Gaussian processes \citep{sauer2023active}, and robust distributed Bayesian inference \citep{andros2024robust}. Full inference typically requires efficient algorithms, such as Markov chain Monte Carlo (MCMC) \citep{guhaniyogi2022distributed,guhaniyogi2023distributed}, variational approximations \citep{gal2016dropout}, and Gaussian Markov random field approximations (\cite{rue2009approximate,lindgren2011explicit} and references therein). Beyond efficient variants of GPs, alternative Bayesian modeling of surrogates have emerged as competitive options. For instance, BASS \citep{francom2020bass} employs adaptive splines for functional data modeling, while BART \citep{chipman2010bart} leverages additive regression trees, and BPPR \citep{collins2024bayesian} uses a Bayesian version of projection pursuit regression.
\cite{higdon2008computer} uses an FPCA approach but used Gaussian processes to model the reduced components. This is also used in \cite{francom2019inferring} with BASS to model the reduced components. Similarly, \cite{bayarri2007comparing} project the functional response onto wavelets and model the reduced components with GPs.
\cite{gu2016parallel} uses a GP to model functional response by building a GP for each output, but sharing some parameters across GPs. \cite{francom2025elastic} treats the functional response misalignment explicitly in the emulation.

In recent years, a rapidly growing area of exploration is the integration of deep neural networks (DNNs) into functional data analysis for rapid computation with big data, primarily in the setting where a functional input is employed in the modeling of a scalar output. For instance, \cite{rossi2002functional, rossi2005functional} were among the first to introduce the concept of a functional neural network (FNN), incorporating functional neurons in the first hidden layer to process functional inputs. This approach was later extended by \cite{thind2020neural, thind2023deep}, enabling FNNs to handle both functional and scalar variables as inputs while modeling a scalar output. \cite{rao2020spatio} further enhanced FNNs by incorporating geographically weighted regression and spatial autoregressive techniques to address regression problems involving spatially correlated functional data. Additionally, \cite{wang2020non} proposed a nonlinear function-on-function regression model using a fully connected NN. In another advancement, \cite{yao2021deep} developed a neural network with a novel basis layer, where the hidden neurons were micro NNs designed to perform parsimonious dimension reduction of functional inputs, leveraging information relevant to a scalar output. \cite{wang2024functional} introduced a functional nonlinear learning framework to effectively represent multivariate functional data in a reduced-dimensional feature space. Meanwhile, \cite{wang2023nonlinear} proposed a nonlinear prediction methodology tailored for functional time series.

\subsection{Proposed Model and Novelty of the Contribution}
The aforementioned prior work has focused on neural networks with functional inputs and scalar outputs. In contrast, this article addresses the more complex task of modeling functional outputs with vector-valued inputs using an interpretable deep learning framework. Specifically, the relationship between the functional output and inputs is represented through an infinite collection of spatially indexed regression functions, each capturing the local association between vector inputs and output at a given spatial location. These functions are further decomposed into two sets of unknown functions: one modeling the nonlinear effects of input variables, and the other capturing spatial dependence across output locations. Both sets of functions are modeled using deep neural networks with dropout. We employ \textit{Monte Carlo (MC) dropout} \citep{gal2016dropout} to generate point predictions and quantify uncertainty in the predicted functional output.

\noindent \underline{\textbf{Novelty of the proposed approach.}} Three main contributions are made in this paper. \textbf{(1) The first Interpretable Bayesian deep neural network architecture in FoS regression.}  Our approach leverages the connection between deep Gaussian processes and dropout-based deep neural networks (DNNs) to generate approximate posterior samples from spatially-indexed regression functions which are modeled with the DNN architecture. These functions encapsulates the relationship between vector-valued inputs and functional outcomes. The resulting samples are then used to obtain posterior predictive draws of the functional response, as described in Section 3.2. This architecture enables rigorous uncertainty quantification for the functional response and enhances interpretability of the surrogate model with DNN architecture, advancing the field of explainable AI for functional data. To our knowledge, this is the first interpretable DNN-based statistical surrogate model for FoS regression with principled uncertainty quantification. There is a recently proposed related article on DNN-based FoS model \cite{wu2023neural} that projects the functional response onto a lower-dimensional space using basis expansions or functional principal components, then models the resulting coefficients using scalar inputs. While effective for point prediction, their method does not quantify uncertainty, a crucial component in statistical surrogate modeling.
\textbf{(1) Data driven basis function estimation.} Much of the functional data literature relies on fixed basis expansions (e.g., B-splines, wavelets) to model unknown regression functions by estimating basis coefficients. In contrast, our approach decomposes the regression function into two sets of unknown functions: one capturing input effects and the other modeling spatial dependence. Estimating both via DNNs offers key advantages, most notably, adaptive modeling of spatial variation and complex nonlinear input-output relationships without the need for manually specified bases or additional structural adjustments. Second, DNNs are inherently optimized to enhance predictive performance and are versatile, enabling them to generalize across a broad range of functional data types and domains. Unlike fixed basis functions, which necessitate extensive manual adjustments, DNN-based modeling provides a highly efficient, task-specific representation, streamlining the modeling process. \textbf{(3) Scalability and inferential accuracy.} Our framework is highly scalable, an end-to-end analysis for 20 simulations and approximately 50,000 spatial locations per simulation within 10 minutes, demonstrating its suitability for large-scale simulations. Its computation time is comparable with state-of-the-art scalable GP methods for FoS regression, while providing significantly improved uncertainty quantification, as demonstrated through empirical evaluations. To our knowledge, our proposed method represents the first interpretable AI framework specifically designed for modeling functional surrogates of computer models, bridging the gap between predictive accuracy, scalability and interpretability.

\section{Model Development}
Let $\bz_h=(z_{h,1},...,z_{h,p})^T$ represent the $p$-dimensional input for the $h$-th simulation, which generates a functional output
$(y_h(\bs):\bs\in\mathcal{S})^T$, where $\mathcal{S}\subseteq\mathbb{R}^2$ denotes a compact domain. Additionally, assume that each functional outcome is influenced at each location $\bs$ by a vector of covariates $\bx(\bs)=(x_1(\bs),..,x_q(\bs))^T$, which is common to all simulations and referred to as \emph{fine-scale covariates}. Our focus is on modeling the nonlinear relationship between the vector-valued input and the functional outcome, while accounting for the effects of the fine-scale covariates. We capture the effects of vector input using a semi-parametric additive regression framework, expressed as,
\begin{align}\label{additive_local_global}
y_h(\bs)=\beta_0+\bx(\bs)^T\bbeta+f_{\bs}(\bz_h)+\epsilon_h(\bs),\:\bs\in\mathcal{S},\:\:h=1,..,H,
\end{align}
where $\beta_0\in\mathbb{R}$ is the intercept, $\bbeta\in\mathbb{R}^q$ is the coefficient vector capturing the effect of fine-scale covariates, and $f_{\bs}(\cdot)$
captures the nonlinear influence of the vector-valued input at location $\bs$. The idiosyncratic errors are assumed to follow
$\epsilon_h(\bs)\stackrel{i.i.d.}{\sim} N(0,\delta^2)$. While this model assumes Gaussian functional outcomes, it can be readily extended to accommodate scenarios where $y_h(\bs)$ is binary, categorical or count-valued. Since our focus is only high-fidelity simulations, we assume the number of simulations $H$ to be moderate.

Let $\mathcal{F}(\mathcal{S})=\{f_{\bs}(\cdot):\bs\in\mathcal{S}\}$ denote the uncountable set of functions, where each function characterizes the effect of inputs at a specific spatial location. Our goal is to jointly model the functions in $\mathcal{F}(\mathcal{S})$ while enabling similar effect of inputs at close-by locations. Specifically, for locations $\bs$ and $\bs'$ that are close to each other in $\mathcal{S}$, the effect of the input $\bz_h$ encoded in functions $f_{\bs}(\cdot)$ and $f_{\bs'}(\cdot)$ will be similar. To achieve this with added flexibility, each function $f_{\bs}(\cdot)$ is modeled with a linear combination of $K$ functions $B_1(\bz_h),...,B_K(\bz_h)$ encoding the effect of inputs with their corresponding spatially-varying coefficient functions $\eta_1(\bs),...,\eta_K(\bs)$, respectively, given by,
\begin{align}\label{basis_expansion}
f_{\bs}(\bz_h)=\sum_{k=1}^K B_k(\bz_h)\eta_k(\bs)=\bet(\bs)^T\bB(\bz_h),\:\:\bs\in\mathcal{S},\:\:h=1,..,H,
\end{align}
where $\bet(\bs)=(\eta_1(\bs),...,\eta_K(\bs))^T$ is the vector of coefficient function at location $\bs$ and $\bB(\bz_h)=(B_1(\bz_h),...,B_K(\bz_h))^T$ is the vector of functions encoding the effect of $\bz_h$.

Although the framework in equation~\ref{basis_expansion} bears a strong resemblance to the basis function representation commonly used for unknown function estimation, where the $B_k(\cdot)$'s serve as basis functions and the $\eta_k(\cdot)$'s as their corresponding coefficients, there are important distinctions that set it apart.
A substantial body of existing literature assumes pre-specified basis functions, with B-spline bases being the most widely used due to their attractive theoretical properties.
Other commonly used options include wavelet bases \citep{vidakovic2009statistical, cressie2015statistics}, radial basis functions \citep{bliznyuk2008bayesian}, and locally bi-square \citep{cressie2008fixed}. However, these pre-defined basis functions often demand substantial manual adjustments to accommodate complex functional data types and domains, especially when addressing nonlinearity in regression \citep{mclean2014functional,wang2016functional,yao2021deep}. Second, in traditional settings, the coefficients of these basis functions are usually treated as scalar quantities, often regularization applied on them to facilitate function estimation. While such approaches are suitable for modeling a single function, our objective is to model an infinite collection of functions indexed by spatial locations. To address this objective, this article treats both $B_k(\cdot)$'s and $\eta_k(\cdot)$'s as unknown and estimates them using deep neural network architectures, providing significantly enhanced flexibility. Details of this approach are outlined in the following section.

\subsection{Deep Neural Network Architecture for Unknown Functions}

We consider a deep neural network architecture with $L_B$ layers, with the $l_B$th layer having $k_{l_B}^{(B)}$ nodes ($l_B=1,\cdots, L_B$), to model the unknown $K$-dimensional vector of functions $\bB(\bz)\in\mathbb{R}^K$ evaluated at an input $\bz$, given by
\begin{align}\label{eq:func_DNN}
    \mathbf{B}(\bz) = \sigma_{L_B}^{(B)} \Big(\mathbf{W}_{L_B}^{(B)} \sigma_{L_{B-1}}^{(B)} \Big( \cdots \sigma_{1}^{(B)} \Big(\mathbf{W}_{1}^{(B)}\bz +\mathbf{b}_{1}^{(B)}\Big)\odot\br^{(B)}_1 \cdots \Big)\odot\br^{(B)}_{L_{B-1}} +\mathbf{b}_{L_B}^{(B)} \Big),
\end{align}
where $\odot$ denotes the element-wise product between two vectors/matrices of the same dimensions and $\br^{(B)}_{1}\in\mathbb{R}^{k^{(B)}_{1}},\ldots,\br^{(B)}_{L_{B-1}}\in\mathbb{R}^{k^{(B)}_{L_{B-1}}}$ are the dropout vectors with binary entries in $\lbrace0,1\rbrace$. The weight matrix $\mathbf{W}^{(B)}_{l_B} \in \mathbb{R}^{k_{l_B}^{(B)} \times k_{l_{B-1}}^{(B)}}$ connects the $(l_{B}-1)$th layer to the $l_B$th hidden layer, $\sigma_{l_B}^{(B)}$ is the activation function for the $l_B$th layer, and $\mathbf{b}^{(B)}_{l_B} \in \mathbb{R}^{k_{l_B}^{(B)}}$ represents the bias parameter.

\begin{figure}[htbp]
\begin{center}
\includegraphics[width = 1\textwidth]{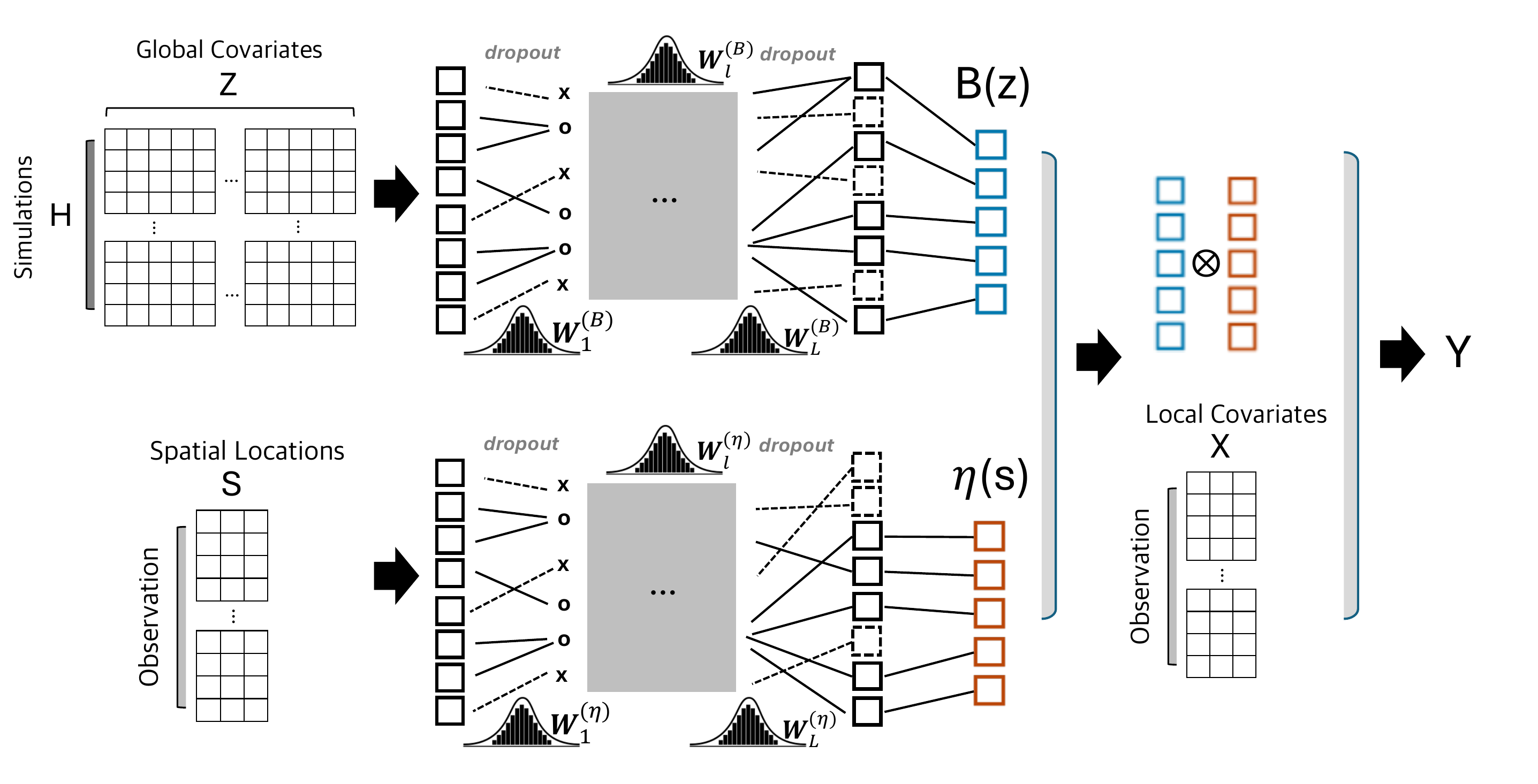}
\end{center}
\caption[]{A graphical illustration of AI-driven DeepSurrogate model}
\label{fig:model}
\end{figure}

Similarly, we adopt a DNN architecture with $L_\eta$ layers to model the vector of coefficient functions, $\bet(\bs)\in \mathbb{R}^{K}$. With the $l_\eta$th layer containing $k_{l_\eta}^{(\eta)}$ nodes, the architecture is specified as,
\begin{align}\label{eq:coef_DNN}
\bet(\bs) = \sigma_{L_\eta}^{(\eta)} \Big(\mathbf{W}_{L_\eta}^{(\eta)} \sigma_{L_{\eta-1}}^{(\eta)} \Big( \cdots \sigma_{1}^{(\eta)} \Big(\mathbf{W}_{1}^{(\eta)}\bs +\mathbf{b}_{1}^{(\eta)}\Big)\odot\br^{(\eta)}_1 \cdots \Big)\odot\br^{(\eta)}_{l_{\eta-1}}  +\mathbf{b}_{L_\eta}^{(\eta)} \Big),
\end{align}
where $\br^{(\eta)}_{1}\in\mathbb{R}^{k^{(\eta)}_{1}},\ldots,\br^{(\eta)}_{L_{\eta-1}}\in\mathbb{R}^{k^{(\eta)}_{L_{\eta-1}}}$ are the dropout vectors with binary entries in $\lbrace0,1\rbrace$.
Here $\mathbf{W}_{l_\eta}^{(\eta)} \in \mathbb{R}^{k_{l_\eta}^{(\eta)} \times k_{l_{\eta-1}}^{(\eta)}}$ is the weight matrix, $\sigma_{l_\eta}^{(\eta)}$ is the activation function for the $l_\eta$th layer, and $\mathbf{b}_{l_\eta}^{(\eta)} \in \mathbb{R}^{k_{l_\eta}^{(\eta)}}$ represents the bias parameter, $l_\eta=1,...,L_\eta$. The DNN architecture in (\ref{eq:coef_DNN}) ensures that if $\bs$ and $\bs'$ are sufficiently close, $\bet(\bs)$ is close to $\bet(\bs')$. Furthermore, the DNN architecture in  (\ref{eq:coef_DNN}) facilitates information sharing between functions in $\mathcal{F}(\mathcal{S})$ across all the spatial locations in $\mathcal{S}$ through the shared weight matrices and bias vectors. Similarly, the DNN architecture of $\bB(\bz)$ in (\ref{eq:func_DNN}) shares information across the input space through its shared weight matrices and bias vectors. By training $\mathbf{B}(\mathbf{z})$ and the coefficient functions $\bet(\mathbf{s})$ across different layers in DNN, as shown in \eqref{eq:func_DNN} and \eqref{eq:coef_DNN}, our model combine them in the final layer of the DNN, along with the local covariate, to construct the model in \eqref{additive_local_global}. The proposed approach AI-driven surrogate modeling approach is referred to as the \emph{DeepSurrogate}. Figure~\ref{fig:model} illustrates the training mechanism for DeepSurrogate.

A key component of our representation is the selection of $K$, which plays the role in choosing the architecture of our deep neural network. In practice, $K$ is treated as part of the model design and is typically selected using standard techniques such as validation performance or cross-validation. Notably, the overall flexibility and expressiveness of the model are determined by the full network architecture, not only the choice of $K$ in the final layer of the neural network. We observed that when the choice of $K$ is made sufficiently large, the resulting inference is less sensitive due to perturbing $K$, as is typically observed in standard basis function representation of functions \citep{guhaniyogi2024bayesian}.

With data observed at locations $\bs_1,...,\bs_n$ for each of the $H$ simulations, a traditional deep neural network approach proceeds to train the model by minimizing the following loss function,
\begin{align}\label{eq:loss}
\mathcal{L}_{\text{DNN}}(\beta_0,\bbeta,\delta^2,\btheta) &:= \frac{1}{2nH} \sum_{i=1}^n\sum_{h=1}^{H}  \frac{(y_{h}(\bs_i) - \widehat{y}_{h}(\bs_i))^2}{\delta^2}
+\sum_{l_B=1}^{L_B} \lambda_{l_B}^{(W)} \left\| \mathbf{W}_{l_B}^{(B)} \right\|_2^2
+\sum_{l_B=1}^{L_B} \lambda_{l_B}^{(b)} \left\| \mathbf{b}_{l_B}^{(B)} \right\|_2^2,
 \nonumber \\
&\qquad\qquad\qquad\qquad\qquad\qquad + \sum_{l_\eta=1}^{L_\eta} \lambda_{l_\eta}^{(W)} \left\| \mathbf{W}_{l_\eta}^{(\eta)} \right\|_2^2
+\sum_{l_\eta=1}^{L_\eta} \lambda_{l_\eta}^{(b)} \left\| \mathbf{b}_{l_\eta}^{(\eta)} \right\|_2^2
\end{align}
where $||\cdot||_2$ denotes the $l_2$ norm, $\widehat{y}_h(\bs_i)=\beta_0+\bx(\bs_i)'\bbeta+\bet(\bs_i)'\bB(\bz_h)$, and $\lambda_{l_B}^{(W)},\lambda_{l_B}^{(b)},\lambda_{l_\eta}^{(W)},\lambda_{l_\eta}^{(b)}$ are penalty parameters controlling shrinkage for weight and bias parameters. The set of unknown parameters $\btheta=\lbrace (\mathbf{W}_{l_B}^{(B)}, \mathbf{b}_{l_B}^{(B)}), l_B=1,\cdots,L_B \rbrace\cup \lbrace (\mathbf{W}_{l_\eta}^{(\eta)}, \mathbf{b}_{l_\eta}^{(\eta)}), l_\eta=1,\cdots,L_\eta \rbrace$ belong to the parameter space $\bTheta$. Equation~\ref{eq:loss} imposes regularization on the connections between layers, effectively minimizing the contribution of interconnections between many pair of neurons in two consecutive layers. This procedure prevents DNN from overfitting, and is popularly known as the \emph{dropout architecture}.
The DNN framework described above provides a data-driven point estimate $\widehat{\btheta}$ for the weight and bias parameters $\btheta$, along with point estimates $\widehat{\beta}_0$, $\widehat{\bbeta}$ and $\widehat{\delta}^2$. However, predictive inference with surrogates requires quantifying predictive uncertainty, naturally achieved through a Bayesian framework by sampling from the approximate posterior distribution of $\btheta$ under an appropriate prior. The next section establishes a key connection between deep Gaussian processes and dropout-based deep neural networks, demonstrating that \emph{the optimization in (\ref{eq:loss}) corresponds to optimizing a specific variational posterior distribution of $\btheta$ corresponding to a specific prior on $\btheta$}. This crucial insight establishes the theoretical backbone of efficient posterior sampling of $\btheta$ by applying random dropout to the estimated $\widehat{\btheta}$ from (\ref{eq:loss}).
This theoretical foundation supports the integration of uncertainty quantification (UQ) in prediction within our framework.

\section{Inference}

\subsection{A Bayesian Deep Neural Network: A Deep Gaussian Process Approximation}

This section introduces an alternative generative framework that applies specific deep Gaussian priors to $\bB(\bz)$ and $\bet(\bs)$, with deep Gaussian process (GP) covariance kernels defined by $\btheta$. We show that a specific variational approximation for the posterior of $\btheta$ converts this into an objective function optimization problem equivalent to the one in (\ref{eq:loss}). This approach builds on the concept proposed by \cite{gal2016dropout}, which demonstrated that applying dropout to each layer of an arbitrarily deep neural network is equivalent to approximating a probabilistic Deep Gaussian Process \citep{damianou2013deep} marginalized over its covariance function. By leveraging this theoretical connection, we can generate Monte Carlo (MC) samples directly from our dropout-enabled DNN architecture, allowing us to accurately quantify predictive uncertainty.

Deep generative models based on deep GPs are first constructed separately for $\bB(\bz)$ and $\bet(\bs)$.
Define a sequence of random linear features $\{\mathbf{f}_{h,l_B}^{(B)}\in\mathbb{R}^{k_{l_B}^{(B)}}: l_B=1,..,L_B;\:h=1,..,H\}$ over $L_B$ layers, with the features in the $l_B$th layer are of dimensions  $k_{l_B}^{(B)}$. Let
$\mathbf{F}_{l_B}^{(B)}=[\mathbf{f}_{1,l_B}^{(B)}:\cdots:\mathbf{f}_{H,l_B}^{(B)}]^T\in\mathbb{R}^{H\times k_{l_B}^{(B)}}$ be a matrix formed by stacking features in the $l_B$th layer, with the $k$th column of $\mathbf{F}_{l_B}^{(B)}$ is given by $\mathbf{f}^{(k)(B)}_{l_B},\:(k=1,\cdots,k_{l_B}^{(B)})$. Similarly, define a sequence of random linear features $\{\mathbf{f}_{i,l_\eta}^{(\eta)}\in\mathbb{R}^{k_{l_\eta}^{(\eta)}}:l_\eta=1,..,L_\eta;\:i=1,..,n\}$ and construct the matrix
$\mathbf{F}_{l_\eta}^{(\eta)}=[\mathbf{f}_{1,l_\eta}^{(\eta)}:\cdots:\mathbf{f}_{n,l_\eta}^{(\eta)}]^T\in\mathbb{R}^{n\times k_{l_\eta}^{(\eta)}}$ by stacking features in the $l_\eta$th layer. Let the $k$th column of $\mathbf{F}_{l_\eta}^{(\eta)}$ be given by $\mathbf{f}^{(k)(\eta)}_{l_\eta},\:(k=1,\cdots,k_{l_\eta}^{(\eta)})$. Assuming independence between columns of $\bF_{l_B}^{(B)}$ given $\bF_{l_{B-1}}^{(B)}$, and independence between columns of $\bF_{l_\eta}^{(\eta)}$ given $\bF_{l_{\eta-1}}^{(\eta)}$, we construct two deep Gaussian processes with $L_B$ and $L_\eta$ layers as a generative models for $\bB(\bz_h)$ and $\bet(\bs_i),$ respectively. Conditional on the previous layer, the finite dimensional realization of the deep GP for the two generative models are given by,
\begin{equation}
\begin{split}
\mathbf{f}^{(k)(B)}_{l_B}|\mathbf{F}^{(B)}_{l_{B-1}}, \bW_{l_B}^{(B)}, \bb_{l_B}^{(B)} &\sim N(\bzero, \mathbf{\Sigma}^{(B)}_{l_B}),\quad l_B=2,\dots,L_B\\
\mathbf{f}^{(k)(\eta)}_{l_\eta}|\mathbf{F}^{(\eta)}_{l_{\eta-1}},\bW_{l_\eta}^{(\eta)}, \bb_{l_\eta}^{(\eta)}  &\sim N(\bzero, \mathbf{\Sigma}^{(\eta)}_{l_\eta}),\quad l_\eta=2,\dots,L_\eta,
\label{eq:fs}
\end{split}
\end{equation}
where the covariance matrices $\bSigma_{l_B}^{(B)}\in \mathbb{R}^{H\times H}$, $\bSigma_{l_\eta}^{(\eta)}\in\mathbb{R}^{n\times n}$ for the finite dimensional realizations depend on the previous layer. Specifically,
\begin{align}\label{covfuncsmultilayer}
&\mathbf{\Sigma}_{l_B}^{(B)}=\int \sigma_{l_B}^{(B)}(\bm\Phi_{l_B-1}^{(B)}\bw^{(B)}+ \mathbf{b}^{(B)})
    \sigma_{l_B}^{(B)}(\bm\Phi_{l_B-1}^{(B)}\bw^{(B)}+ \mathbf{b}^{(B)})^{\top}p(\bw^{(B)})p(\mathbf{b}^{(B)})d\bw^{(B)}d\mathbf{b}^{(B)},\nonumber\\
    &\mathbf{\Sigma}_{l_\eta}^{(\eta)}=\int \sigma_{l_\eta}^{(\eta)}(\bm\Phi_{l_\eta-1}^{(\eta)}\bw^{(\eta)}+ \mathbf{b}^{(\eta)})
    \sigma_{l_\eta}^{(\eta)}(\bm\Phi_{l_\eta-1}^{(\eta)}\bw^{(\eta)}+ \mathbf{b}^{(\eta)})^{\top}p(\bw^{(\eta)})p(\mathbf{b}^{(\eta)})d\bw^{(\eta)}d\mathbf{b}^{(\eta)},
\end{align}
where $\bPhi_{l_B}^{(B)}=[\bphi_{1,l_B}^{(B)}:\cdots:\bphi_{H,l_B}^{(B)}]^T\in\mathbb{R}^{H\times k^{(B)}_{l_B}}$ are the collection of feature vectors after applying the activation function, such that $\bphi_{h,l_B}^{(B)}=\sigma_{l_B}^{(B)}(\mathbf{f}_{h,l_B}^{(B)})$. Similarly, $\bPhi_{l_\eta}^{(\eta)}=[\bphi_{1,l_\eta}^{(\eta)}:\cdots:\bphi_{n,l_\eta}^{(\eta)}]^T\in\mathbb{R}^{n\times k^{(\eta)}_{l_\eta}}$ are the collection of feature vectors after transformation such that $\bphi_{i,l_\eta}^{(\eta)}=\sigma_{l_\eta}^{(\eta)}(\mathbf{f}_{i,l_\eta}^{(\eta)})$.

Approximating the expectations in \eqref{covfuncsmultilayer} by averaging over the $k_{l_B}^{(B)}$ hidden-unit weight vectors, corresponding to the rows of $\bW_{l_B}^{(B)}$ at the $l_B$th layer, and
$\bSigma^{(\eta)}_{l_\eta}$ through MC integration over $k_{l_\eta}^{(\eta)}$ rows of $\bW_{l_\eta}^{(\eta)}$ yield,
\begin{align}
\widehat{\mathbf{\Sigma}}_{l_B}^{(B)} &= \frac{1}{k_{l_B}^{(B)}}\sigma_{l_B}(\bm\Phi_{l_B-1}^{(B)}\mathbf{W}^{\top (B)}_{l_B}+ \mathbf{1}_H\otimes\mathbf{b}_{l_B}^{(B)})
    \sigma_{l_B}(\bm\Phi_{l_B-1}^{(B)}\mathbf{W}^{\top(B)}_{l_B}+ \mathbf{1}_H\otimes\mathbf{b}_{l_B}^{(B)})^{\top}\nonumber\\
\widehat{\mathbf{\Sigma}}_{l_\eta}^{(\eta)} &= \frac{1}{k_{l_\eta}^{(\eta)}}\sigma_{l_\eta}(\bm\Phi_{l_\eta-1}^{(\eta)}\mathbf{W}^{\top(\eta)}_{l_\eta}+ \mathbf{1}_n\otimes\mathbf{b}_{l_\eta}^{(\eta)})
    \sigma_{l_\eta}(\bm\Phi_{l_\eta-1}^{(\eta)}\mathbf{W}^{\top(\eta)}_{l_\eta}+ \mathbf{1}_n\otimes\mathbf{b}_{l_\eta}^{(\eta)})^{\top},
    \label{covfuncsmultilayer_mc}
\end{align}
where \(\mathbf{1}_H\) and  \(\mathbf{1}_n\) denote vectors of ones of lengths $H$ and $n$, respectively, used to replicate the bias terms across simulations and spatial locations. Here \eqref{covfuncsmultilayer_mc} becomes accurate in approximating the true covariance matrix as $k_{l_B}^{(B)}$ and $k_{l_\eta}^{(\eta)}$ increase. Assuming that the final layer in both generative models consists of $K$ components, i.e., $k_{L_B}^{(B)}=k_{L_\eta}^{(\eta)}=K$, we estimate the functions $\bB(\bz)$ and the corresponding coefficients $\bet(\bs)$ using the features from this last layer. Specifically, we set
 $\bB(\bz_h)= {\boldsymbol f}_{h,L_B}^{(B)}$ ($i=1,...,n$) and $\bet(\bs_i)={\boldsymbol f}_{i,L_\eta}^{(\eta)}$ ($i=1,...,n$). The generative framework for the DeepSurrogate is given by,
\begin{align}\label{eq:fs_last}
& y_h(\bs_i)|\mathbf{F}^{(B)}_{L_B}, \mathbf{F}^{(\eta)}_{L_\eta}, \bx(\bs_i) \sim N(y_h(\bs_i)|\beta_0+\bx(\bs_i)^T\bbeta+\bB(\bz_h)^T\bet(\bs_i),\delta^2),\nonumber\\
&\qquad\qquad\bB(\bz_h)={\boldsymbol f}_{h,L_B}^{(B)},\:\bet(\bs_i)={\boldsymbol f}_{i,L_\eta}^{(\eta)},\nonumber\\
&\qquad\qquad\mathbf{f}^{(k)(B)}_{l_B}|\mathbf{F}^{(B)}_{l_{B-1}}, \bW_{l_B}^{(B)}, \bb_{l_B}^{(B)} \sim N(\bzero, \mathbf{\Sigma}^{(B)}_{l_B}),\quad l_B=2,\dots,L_B;\:k=1,...,k_{l_B}^{(B)}\nonumber\\
&\qquad\qquad\mathbf{f}^{(k)(\eta)}_{l_\eta}|\mathbf{F}^{(\eta)}_{l_{\eta-1}},\bW_{l_\eta}^{(\eta)}, \bb_{l_\eta}^{(\eta)}  \sim N(\bzero, \mathbf{\Sigma}^{(\eta)}_{l_\eta}),\quad l_\eta=2,\dots,L_\eta;\:k=1,...,k_{l_\eta}^{(\eta)}.
\end{align}

Next, we construct a specific variational approximation of the posterior distribution of the weight and bias parameters.
Let $\bw^{(B)}_{l_B,k} = (w^{(B)}_{l_B,k1}, \ldots, w^{(B)}_{l_B,k,k^{(B)}_{l_B-1}})$ and
$\bw^{(\eta)}_{l_\eta,k} = (w^{(\eta)}_{l_\eta,k1}, \ldots, w^{(\eta)}_{l_\eta,k,k^{(\eta)}_{l_\eta-1}})$ denote the $k$th row of the weight matrices $\bW^{(B)}_{l_B}$ and $\bW^{(\eta)}_{l_\eta}$, respectively. We define the variational distribution as $\prod_{l_B=1}^{L_B}q(\mathbf{W}^{(B)}_{l_B})q(\mathbf{b}^{(B)}_{l_B})\prod_{l_\eta=1}^{L_\eta}q(\mathbf{W}^{(\eta)}_{l_\eta})q(\mathbf{b}^{(\eta)}_{l_\eta})$ where each factor is further decomposed row-wise variational distribution as
\begin{equation}\begin{split}
    q(\mathbf{W}^{(B)}_{l_B}) & = \prod_{\forall k} q(\bw^{(B)}_{l_B,k}),~~q(\mathbf{b}^{(B)}_{l_B}) = \prod_{\forall k} q(b^{(B)}_{l_B,k}), ~~~  q(\mathbf{W}^{(\eta)}_{l_\eta})=\prod_{\forall k} q(\bw^{(\eta)}_{l_\eta,k}),~~q(\mathbf{b}^{(\eta)}_{l_\eta}) = \prod_{\forall k} q(b^{(\eta)}_{l_\eta,k}),\\
    q(\bw^{(B)}_{l_B,k}) &= p_{l_B}N(\bmu^{w^{(B)}}_{l_B,k},\sigma^2)+(1-p_{l_B})N(\mathbf{0},\sigma^2), ~~ q(\bw^{(\eta)}_{l_\eta,k}) = p_{l_\eta}N(\mu^{w^{(\eta)}}_{l_\eta,k},\sigma^2)+(1-p_{l_\eta})N(0,\sigma^2) \\
    q(b^{(B)}_{l_B,k}) &= p_{l_B}N(\bmu^{b^{(B)}}_{l_B,k},\sigma^2)+ (1-p_{l_B})N(\mathbf{0},\sigma^2) ~~  q(b^{(\eta)}_{l_\eta,k}) = p_{l_\eta}N(\mu^{b^{(\eta)}}_{l_\eta,k},\sigma^2)+ (1-p_{l_\eta})N(0,\sigma^2) ,
\end{split}
\label{variationaldist}
\end{equation}
where $l_B = 1, \ldots, L_B$ and $l_\eta = 1, \ldots, L_\eta$. Here ${\bmu}^{w^{(B)}}_{l_B,k} \in \mathbb{R}^{k^{(B)}_{l_B - 1}}$ and
${\bmu}^{w^{(\eta)}}_{l_\eta,k} \in \mathbb{R}^{k^{(\eta)}_{l_\eta - 1}}$
are the variational mean vectors for the $k$th rows of $\bW^{(B)}_{l_B}$ and $\bW^{(\eta)}_{l_\eta}$, respectively,
and $\mu^{b^{(B)}}_{l_B,k}$, $\mu^{b^{(\eta)}}_{l_\eta,k}$ are the variational means for the corresponding bias elements. Each weight or bias parameter follows a two-component Gaussian mixture sharing a common small variance $\sigma^2$.

Let $\tilde{\btheta}$ denote the collection of all variational parameters $\{{\bmu}^{w^{(B)}}_{l_B,k} : l_B = 1,\ldots,L_B;\; k = 1,\ldots,k^{(B)}_{l_B}\}$,
$\{{\bmu}^{w^{(\eta)}}_{l_\eta,k} : l_\eta = 1,\ldots,L_\eta;\; k = 1,\ldots,k^{(\eta)}_{l_\eta}\}$,
$\{\mu^{b^{(B)}}_{l_B,k} : l_B = 1,\ldots,L_B;\; k = 1,\ldots,k^{(B)}_{l_B}\}$,
$\{\mu^{b^{(\eta)}}_{l_\eta,k} : l_\eta = 1,\ldots,L_\eta;\; k = 1,\ldots,k^{(\eta)}_{l_\eta}\}$,
and $\sigma^2$. As the inclusion probabilities $p_{l_B}, p_{l_\eta} \in [0,1]$ approach~$0$, $q(\bw^{(B)}_{l_B,k}),q( b^{(B)}_{l_B,k}),q(\bw^{(\eta)}_{l_\eta,k}),$ and $q(b^{(\eta)}_{l_\eta,k})$
become $\mathcal{N}(\mathbf{0}, \sigma^2\bI)$, indicating that the corresponding parameters are
effectively dropped (i.e., set to zero).
We denote the full variational distribution as $q(\btheta \mid \tilde{\btheta})$ to make its dependence
on the variational parameters explicit.
The optimal variational parameters are obtained by maximizing the evidence lower bound (ELBO),
\begin{equation}
  \mathbb{E}_{q}\bigl[\log \pi(\btheta, \bD)\bigr]
  - \mathbb{E}_{q}\bigl[\log q(\btheta \mid \tilde{\btheta})\bigr],
\label{eq:elbo}
\end{equation}
where $\pi(\btheta, \bD)$ denotes the joint distribution of the data $\bD$ and parameters $\btheta$.

Let $\btheta^{(m)}$ be the $m$th Monte Carlo~(MC) sample drawn from $q(\btheta\mid\tilde{\btheta})$ then the MC approximation of the ELBO is given by
\begin{align}
  \mathcal{L}_{\mathrm{MC}}(\sigma^2,\, \tilde{\btheta})
  &\approx \frac{1}{M}\sum_{m=1}^{M} \sum_{h=1}^{H} \sum_{i=1}^{n}
    \log p\!\left(\by_h(\bs_i) \mid \bz_h, \bx(\bs_i),\, \btheta^{(m)},\beta_0,\bbeta, \delta^2 \right) \nonumber\\
  &\quad - \sum_{l_B=1}^{L_B} \frac{p_{l_B}}{2}
    \left(\bigl\|\boldsymbol{\mu}^{w^{(B)}}_{l_B}\bigr\|_2^2
         + \bigl\|\boldsymbol{\mu}^{b^{(B)}}_{l_B}\bigr\|_2^2\right)
  - \sum_{l_\eta=1}^{L_\eta} \frac{p_{l_\eta}}{2}
    \left(\bigl\|\boldsymbol{\mu}^{w^{(\eta)}}_{l_\eta}\bigr\|_2^2
         + \bigl\|\boldsymbol{\mu}^{b^{(\eta)}}_{l_\eta}\bigr\|_2^2\right),
\label{eq:mc_elbo}
\end{align}
where $\boldsymbol{\mu}^{w^{(B)}}_{l_B}$ and $\boldsymbol{\mu}^{b^{(B)}}_{l_B}$ collect all variational
means of the weights and biases in layer $l_B$, and analogously for $l_\eta$.

\subsection{Approximate Posterior Sampling of DNN using MC Dropout}

By scaling the negative of $\mathcal{L}_{\mathrm{MC}}$ in
\eqref{eq:mc_elbo} by $1/(nH)$ and setting the regularization coefficients in \eqref{eq:loss} $\lambda_{l_B}^{(W)} = \lambda_{l_B}^{(b)} = \dfrac{p_{l_B}}{2nH}$ and $\lambda_{l_\eta}^{(W)} = \lambda_{l_\eta}^{(b)} = \dfrac{p_{l_\eta}}{2nH}$, the negative scaled \eqref{eq:mc_elbo} becomes equivalent to the training loss function with $\ell_2$ in \eqref{eq:loss}. This reveals that minimizing the regularized training loss to obtain $\widehat{\btheta}$ is mathematically equivalent to maximizing the ELBO to obtain optimal variational parameters $\tilde{\btheta}$. This equivalence is consequential: it permits efficient estimation of the variational parameters via stochastic gradient descent~(SGD) applied to the training loss.

Assuming $\sigma^2 \approx 0$, approximate posterior samples of $\btheta$ can then be generated by applying dropout to $\widehat{\btheta}$.
Specifically, for each layer $l_B = 1,\ldots,L_{B}-1$, we generate a vector of Bernoulli random variables $\br^{(B)}_{l_B} \in \{0,1\}^{k^{(B)}_{l_B}}$ with keep probability $p_{l_B}$,
and construct the dropout mask
\[
  \bR^{(B)}_{l_B}
  = \br^{(B)}_{l_B}\,\mathbf{1}^\top_{k^{(B)}_{l_B-1}}
  \in \mathbb{R}^{k^{(B)}_{l_B} \times k^{(B)}_{l_B-1}}.
\]
Similarly, for each layer $l_\eta = 1,\ldots,L_{\eta}-1$, we generate $\br^{(\eta)}_{l_\eta} \in \{0,1\}^{k^{(\eta)}_{l_\eta}}$ with keep probability $p_{l_\eta}$,
and construct
\[
  \bR^{(\eta)}_{l_\eta}
  = \br^{(\eta)}_{l_\eta}\,\mathbf{1}^\top_{k^{(\eta)}_{l_\eta-1}}
  \in \mathbb{R}^{k^{(\eta)}_{l_\eta} \times k^{(\eta)}_{l_\eta-1}}.
\]
Applying these masks element-wise to the estimated parameters yields
\[
  \widehat{\bW}^{(B)}_{l_B} \odot \bR^{(B)}_{l_B}, \quad
  \widehat{\bb}^{(B)}_{l_B} \odot \br^{(B)}_{l_B}, \quad
  \widehat{\bW}^{(\eta)}_{l_\eta} \odot \bR^{(\eta)}_{l_\eta}, \quad
  \widehat{\bb}^{(\eta)}_{l_\eta} \odot \br^{(\eta)}_{l_\eta},
\]
which, under $\sigma^2 \approx 0$, constitute approximate draws from the marginal variational posteriors of $\bW^{(B)}_{l_B}$, $\bb^{(B)}_{l_B}$, $\bW^{(\eta)}_{l_\eta}$, and $\bb^{(\eta)}_{l_\eta}$
in~\eqref{variationaldist}, respectively. The optimal values of $p_{l_B}$, $p_{l_\eta}$, and $\sigma^2$ are selected via a grid search. Repeating this procedure $F$ times yields $F$ approximate posterior samples from $\btheta$, which are further used for posterior predictive inference. The algorithm is described in Algorithm 1.

\begin{algorithm}[tbp]
\small
\setlength{\abovedisplayskip}{3pt}
\setlength{\belowdisplayskip}{3pt}
\setlength{\abovedisplayshortskip}{2pt}
\setlength{\belowdisplayshortskip}{2pt}
\caption{DeepSurrogate Algorithm}
\label{alg1}

\noindent\textbf{Input:} Training data $\{y_h(\bs_{i}), \bs_{i}, \bx(\bs_{i}), \bz_{h}\}$.\par\medskip
\noindent\textbf{Output:} Posterior samples of $\btheta$ and $\delta^2$.\par\medskip

\noindent\textbf{Step 1:} Initialize $\btheta^{(0)}$, $\bbeta^{(0)}$, $\beta_0^{(0)}$\par\smallskip

\vspace{0.2em}
\noindent\textbf{Step 2:} Optimize $\btheta$, $\bbeta$, $\beta_0$ via SGD on loss Eq.~\eqref{eq:loss} for $n_{\text{iter}}$ iterations\par\smallskip
\textbf{Return:} $\widehat{\btheta} = \bigl\{ \{\widehat{\mathbf{W}}_{l_B}^{(B)},\, \widehat{\mathbf{b}}_{l_B}^{(B)}\}_{l_B=1}^{L_B},\; \{\widehat{\mathbf{W}}_{l_\eta}^{(\eta)},\, \widehat{\mathbf{b}}_{l_\eta}^{(\eta)}\}_{l_\eta=1}^{L_\eta} \bigr\}$, $\widehat{\beta}_0$, $\widehat{\bbeta}$\par\smallskip

\vspace{0.2em}
\noindent\textbf{Step 3:} Draw posterior samples of $\btheta$ and $\delta^2$ via dropout\par\smallskip
\noindent\textbf{For $f=1,\ldots,F$:}
\begin{quote}
    \textbf{(a) Sample dropout masks:} For each $l_B$, draw
    $\br_{l_B}^{(B)(f)}\sim \mathrm{Bernoulli}(p_{l_B})$ and set $\bR_{l_B}^{(B)(f)}=\br_{l_B}^{(B)(f)}\mathbf{1}_{k_{l_B-1}^{(B)}}^{\top}$; similarly for each $l_\eta$, draw $\br_{l_\eta}^{(\eta)(f)} \sim \mathrm{Bernoulli}(p_{l_\eta})$ and set $\bR_{l_\eta}^{(\eta)(f)} = \br_{l_\eta}^{(\eta)(f)}\mathbf{1}_{k_{l_\eta-1}^{(\eta)}}^{\top}$\par\smallskip

    \textbf{(b) Apply masks:} Set
\[
\btheta^{(f)}=\left\{
\begin{aligned}
&\left\{\widehat W_{l_B}^{(B)}\odot R_{l_B}^{(B)(f)},
\widehat b_{l_B}^{(B)}\odot \boldsymbol r_{l_B}^{(B)(f)}\right\}_{l_B=1}^{L_B},\\
&\left\{\widehat W_{l_\eta}^{(\eta)}\odot R_{l_\eta}^{(\eta)(f)},
\widehat b_{l_\eta}^{(\eta)}\odot \boldsymbol r_{l_\eta}^{(\eta)(f)}\right\}_{l_\eta=1}^{L_\eta}
\end{aligned}
\right\}.
\]

    \textbf{(c) Compute} $\bB^{(f)}(\bz)$ via Eq.~\eqref{eq:func_DNN} and $\bet^{(f)}(\bs)$ via Eq.~\eqref{eq:coef_DNN}\par\smallskip

    \textbf{(d) Compute} $\delta^{2(f)} = \dfrac{1}{nH}\displaystyle\sum_{i=1}^n\sum_{h=1}^H \left(y_h(\bs_i) - \widehat{\beta}_0 - \bx(\bs_i)^\top\widehat{\bbeta} - \bet^{(f)}(\bs_i)^{\top}\bB^{(f)}(\bz_h)\right)^2$\par\smallskip
\end{quote}
\noindent\textbf{End for.}\par

\vspace{0.2em}
\noindent\textbf{Step 4:} Return approximate posterior samples: $\{\btheta^{(f)}\}_{f=1}^F$, $\{\delta^{2(f)}\}_{f=1}^F$.
\end{algorithm}

\subsection{Predictive Distribution}
Let $\bD=\{(\by_h(\bs_i),\bz_h^T,\bx(\bs_i)^T):h=1,..,H;\:i=1,..,n\}$ denotes the observed data.
Suppose the goal is to predict the unobserved response $y^*$ at location $\bs^*$ with the global covariate
$\bz^*$ and local covariate $\bx(\bs^*)$.
The predictive distribution can be constructed as
\begin{equation}\label{predictdist1}
\begin{aligned}
p(y^*|\bD,\bz^*,\bx(\bs^*))
&=\int\cdots\int p(y^{\ast}|\btheta,\beta_0,\bbeta,\delta^2,\bz^*,\bx(\bs^*))\\
&\qquad\times p(\btheta,\beta_0,\bbeta,\delta^2|\bD)\,d\btheta\,d\beta_0\,d\bbeta\,d\delta^2.
\end{aligned}
\end{equation}
where $p(\btheta,\beta_0,\bbeta,\delta^2|\bD)$ denotes the posterior distribution of the parameters.
Since it is challenging to compute \eqref{predictdist1} directly, we approximate it through samples of $\btheta$ drawn using MC dropout architecture discussed in Section 3.1 and Algorithm 1. Specifically, with approximate posterior samples $\{\btheta^{(f)}\}_{f=1}^F=\lbrace\lbrace \mathbf{W}_{l_B}^{(B)(f)}, \mathbf{b}_{l_B}^{(B)(f)} \rbrace_{l_B=1}^{L_B},\lbrace \mathbf{W}_{l_\eta}^{(\eta)(f)}, \mathbf{b}_{l_\eta}^{(\eta)(f)} \rbrace_{l_\eta=1}^{L_\eta}\rbrace_{f=1}^{F}$ and $\{\delta^{2(f)}\}_{f=1}^F$ drawn through Algorithm 1, we adopt composition sampling, wherein the $f$th post burn-in iterate
$y^{*(f)}$ is drawn from $p(y^{\ast (f)}|\bz^*,\bx(\bs^*),\btheta^{(f)},\widehat{\beta}_0,\widehat{\bbeta},\delta^{2(f)})$, given by
 $p(y^{\ast (f)}|\bz^*,\bx(\bs^*),\btheta^{(f)},\widehat{\beta}_0,\widehat{\bbeta},\delta^{2(f)})=N(\widehat{\beta}_0+\bx(\bs^*)^T\widehat{\bbeta}+\bet^{(f)}(\bs^*)^T\bB^{(f)}(\bz^*),\delta^{2(f)}).$
Point prediction and 95\% predictive intervals are computed from these predictive samples.

\section{Empirical Study with Synthetic Data}
\noindent This section compares DeepSurrogate to standard benchmarks for predicting functional outcomes at new simulations using input and fine-scale covariates. All evaluations use a small to moderate number of simulation runs, reflecting practical limits of high-cost simulations.

\subsection{Data Generation}
To simulate the data, we draw $n$
spatial locations $\bs_1,\ldots,\bs_{n}$ uniformly over the domain
$\mathcal{D} = [0,10]\times [0,10]$. The data generation process involves
$p=5$ input variables and $q=2$ fine-scale covariate. To mimic the real data structure, input vector $\bz_h$ for each simulation is generated from a multivariate normal distribution with mean zero and a covariance matrix $\bSigma$, where the off-diagonal elements are set to a common correlation coefficient $\rho=0.1$ and the diagonal elements are set to 1, i.e., $\bz_1,...,\bz_{H+H_0}\stackrel{i.i.d.}{\sim} N(0,\bSigma)$. Similarly, the local attributes $\bx(\bs_1),...,\bx(\bs_{n})$ are simulated independently from $N(\bzero,\bI_q)$. For each $h=1,...,H+H_0$ and $i=1,...,n$, the response
$y_h(\bs_i)$ is drawn independently from $N(\beta_0^*+\bx(\bs_i)^T\bbeta^*+\bet^*(\bs_i)^T\bB(\bz_h),\delta^{*2})$ following (\ref{additive_local_global}) and (\ref{basis_expansion}).

The true intercept $\beta_0^*$ is set at $0.5$, while each component of the coefficient $\bbeta^*$ for fine-scale covariates is drawn from $U(-1.5,1.5)$. The functions $B_k(\bz)$ are chosen as the tensor-product of B-splines of order $4$ with $5$ knots. The true coefficient function $\eta_k(\cdot)$ is simulated from a Gaussian process with mean $0$ and exponential covariance kernel, characterized by the spatial variance $\alpha_k^{*2}$ and scale parameter $l_k^*$, such that Cov($\eta_k(\bs_i),\eta_k(\bs_j))=\alpha_k^{*2}\exp(-||\bs_i-\bs_j||/l_k^*)$.

Out of $H+H_0$ simulations, we randomly choose $H$ simulations for fitting, and the remaining $H_0$ simulations are used for testing the predictive performance of the competitors. We consider a few different scenarios by varying the variance parameter $\alpha_k^*$, scale parameter $l_k^*$, sample size $n$, number of fitted simulations $H$ and noise variance $\delta^{*2}$, as discussed below.

\begin{itemize}
    \item \textbf{Scenario 1}: $n=600$, $H=100$, $H_0=20$, $\alpha_k^{*2}$ are drawn uniformly between $[5, 10]$, and $l_k^*$ is drawn uniformly between $[4, 8]$, for $k=1,..,K$, and $\delta^{*2}=1$.
    \item \textbf{Scenario 2}: $n=600$, $H=100$, $H_0=20$, $\alpha_k^{*2}$ are drawn uniformly between $[0.5, 1]$, and $l_k^*$ is drawn uniformly between $[4, 8]$, for $k=1,..,K$, and $\delta^{*2}=1$.
    \item \textbf{Scenario 3}: $n=600$, $H=100$, $H_0=20$, $\alpha_k^{*2}$ are drawn uniformly between $[0.5, 1]$, and $l_k^*$ is drawn uniformly between $[0.5, 1]$, for $k=1,..,K$, and $\delta^{*2}=1$.
    \item \textbf{Scenario 4}: $n=6000$, $H=20$, $H_0=20$, $\alpha_k^{*2}$ are drawn uniformly between $[5, 10]$, and $l_k^*$ is drawn uniformly between $[4, 8]$, for $k=1,..,K$, and $\delta^{*2}=1$.
    \item \textbf{Scenario 5}: $n=6000$, $H=10$, $H_0=20$, $\alpha_k^{*2}$ are drawn uniformly between $[5, 10]$, and $l_k^*$ is drawn uniformly between $[4, 8]$, for $k=1,..,K$, and $\delta^{*2}=1$.
    \item \textbf{Scenario 6}: $n=6000$, $H=10$, $H_0=20$,  $\alpha_k^{*2}$ are drawn uniformly between $[5, 10]$, and $l_k^*$ is drawn uniformly between $[4, 8]$, for $k=1,..,K$, and $\delta^{*2}=0.5$.
    \item \textbf{Scenario 7}: $n=6000$, $H=10$, $H_0=20$, $\alpha_k^{*2}$ are drawn uniformly between $[5, 10]$, and $l_k^*$ is drawn uniformly between $[4, 8]$, for $k=1,..,K$, and $\delta^{*2}=0.1$.
\end{itemize}
Note that Scenarios 1--5 exhibit a signal-to-noise ratio (SNR) of less than 2, Scenario 6 features an SNR of 3.28, and Scenario 7 has an SNR greater than 5. Higher values of $l_k^*$ correspond to stronger spatial dependence. We also vary the sample size $n$ to assess DeepSurrogate's performance under different data availability, given deep learning's data-intensive nature. We also assess performance across moderate to small numbers of fitted simulations $H$, reflecting the surrogate's use for expensive simulations.

\begin{table} [h!]
\small
\setlength{\tabcolsep}{4pt}
\centering
\begin{tabular}{l l c c c c}
&  &   \textbf{DeepSurrogate} &\textbf{BassPCA} &\textbf{FOSR} &\textbf{VecchiaGP}  \\
& &  &  & \\
\hline
\hline
\textbf{Scenario 1} & RMSPE & 2.0447 & \textbf{2.0244} & 2.4109 & 2.0661 \\
& Time (min) &  4.7651  & 4.8776 & 0.0356 & 0.4922\\
\hline
\textbf{Scenario 2} & RMSPE & \textbf{1.9960} & 2.0118 &  2.0119 & 2.0654 \\
& Time (min) &  4.6387  & 5.2837 & 0.0358 & 0.5156\\
\hline
\textbf{Scenario 3} & RMSPE & 2.0288 &  \textbf{2.0067} &  2.0517 & 2.0656 \\
& Time (min) &  4.2577  & 4.1999  & 0.0447 & 0.6303 \\
\hline
\textbf{Scenario 4} & RMSPE & \textbf{2.1439} & 2.6339 &  2.7022  & 2.1792 \\
& Time (min) &  3.0667  & 5.2639 & 0.0993  & 0.0189\\
\hline
\textbf{Scenario 5} & RMSPE & \textbf{2.2106} & 2.5486 & 2.9741 & 2.2623  \\
& Time (min) &  4.5025  &  4.4748  & 0.0335 & 0.3193 \\
\hline
\textbf{Scenario 6} & RMSPE & \textbf{0.8064} & 1.1112 & 1.5400 & 0.8972  \\

& Time (min) & 5.9694 &  2.1700  & 0.1362 & 1.4918 \\
\hline
\textbf{Scenario 7} & RMSPE & \textbf{0.3660}  & 0.3775 & 1.5951 & 0.7369  \\

& Time (min) & 4.7135&  1.0261  & 0.1672 & 0.0285 \\
\hline
\hline
\end{tabular}
\caption[]{Predictive point estimates for DeepSurrogate and competing methods (BassPCA, FOSR, and VecchiaGP) across seven simulation scenarios are presented. For each method, root mean squared prediction error (RMSPE) and computing time (in minutes) are reported. The best performing model in terms of point prediction is boldfaced under each scenario.}\label{table_simul_spatial}
\end{table}

\subsection{Competitors and Metric of Comparison}
We implement our method using {\tt TensorFlow}, an open-source machine learning platform, and use the \texttt{ReLU} activation function in all hidden layers. Computations were performed on an 8-core AMD Radeon Pro 5500 XT processor. The objective function in (\ref{eq:loss}) is optimized using the \texttt{Adam} optimizer with an initial learning rate of $10^{-2}$, a decay rate of $0.97$, and decay steps of 10,000. We train the model for $500$ epochs with a batch size of $128$.
For all simulations, we compare the proposed DeepSurrogate with function-on-scalar regression (FOSR) \citep{reiss2010fast}, Bayesian adaptive smoothing spline with principal component analysis (BassPCA) \citep{francom2020bass}, a method that combines principal component analysis (PCA) with Bayesian adaptive smoothing spline priors for scalable functional regression, and computationally efficient alternatives for Gaussian process regression by VecchiaGP \citep{katzfuss2021general}. Both VecchiaGP and BassPCA are state-of-the-art in the surrogate literature and are widely used in national laboratories. On the other hand, FOSR is a popular method used in the functional data analysis.

The point prediction accuracy of the competitors is evaluated using mean squared prediction error (MSPE), which measures the discrepancy between the true and the predicted responses
$\sum_{h=H+1}^{H+H_0}\sum_{i=1}^n(y_h(\bs_i)-\widehat{y}_h(\bs_i))^2/(nH_0)$, where $\widehat{y}_h(\bs_i)$ denotes the point prediction of
$y_h(\bs_i)$, which is taken to be the posterior mean for Bayesian competitors. The predictive uncertainty is determined with the coverage probability and interval length for 95\% predictive intervals (PIs) for all the competing methods
over $H_0$ out of sample simulations.

\subsection{Results}
Table~\ref{table_simul_spatial} summarizes the point prediction performance across competing methods. All models demonstrate strong performance in Scenarios 1--3, where abundant functional observations are available, and their accuracy remains stable across varying levels of spatial correlation and spatial variance. However, when the number of simulations is limited and SNR is low, as in Scenarios 4 and 5, DeepSurrogate and VecchiaGP consistently achieve superior point prediction accuracy by effectively utilizing spatial associations to model the spatially varying relationship between inputs and outputs, even with fewer observations. In contrast, BassPCA and FOSR, which rely on basis function expansions to represent functional responses, suffer from instability in parameter estimation under limited sample sizes, as the number of parameters can significantly exceed the number of available data points, leading to poorer predictive performance. Further, Scenarios 6 and 7, representing limited simulations with higher SNR, demonstrate DeepSurrogate's significantly superior performance over all other competitors.
Since computer simulation models often produce data with relatively high SNR---as also demonstrated in Section 5---DeepSurrogate appears particularly well-suited for such settings. This is supported by its strong performance in Scenarios 6 and 7, as shown in Table~\ref{table_simul_spatial}. All methods are computationally efficient, with DeepSurrogate completing full predictive inference across 20 simulations, each involving 6,000 observations, in under five minutes.

\begin{figure}[h]
\centering\includegraphics[width=\linewidth]{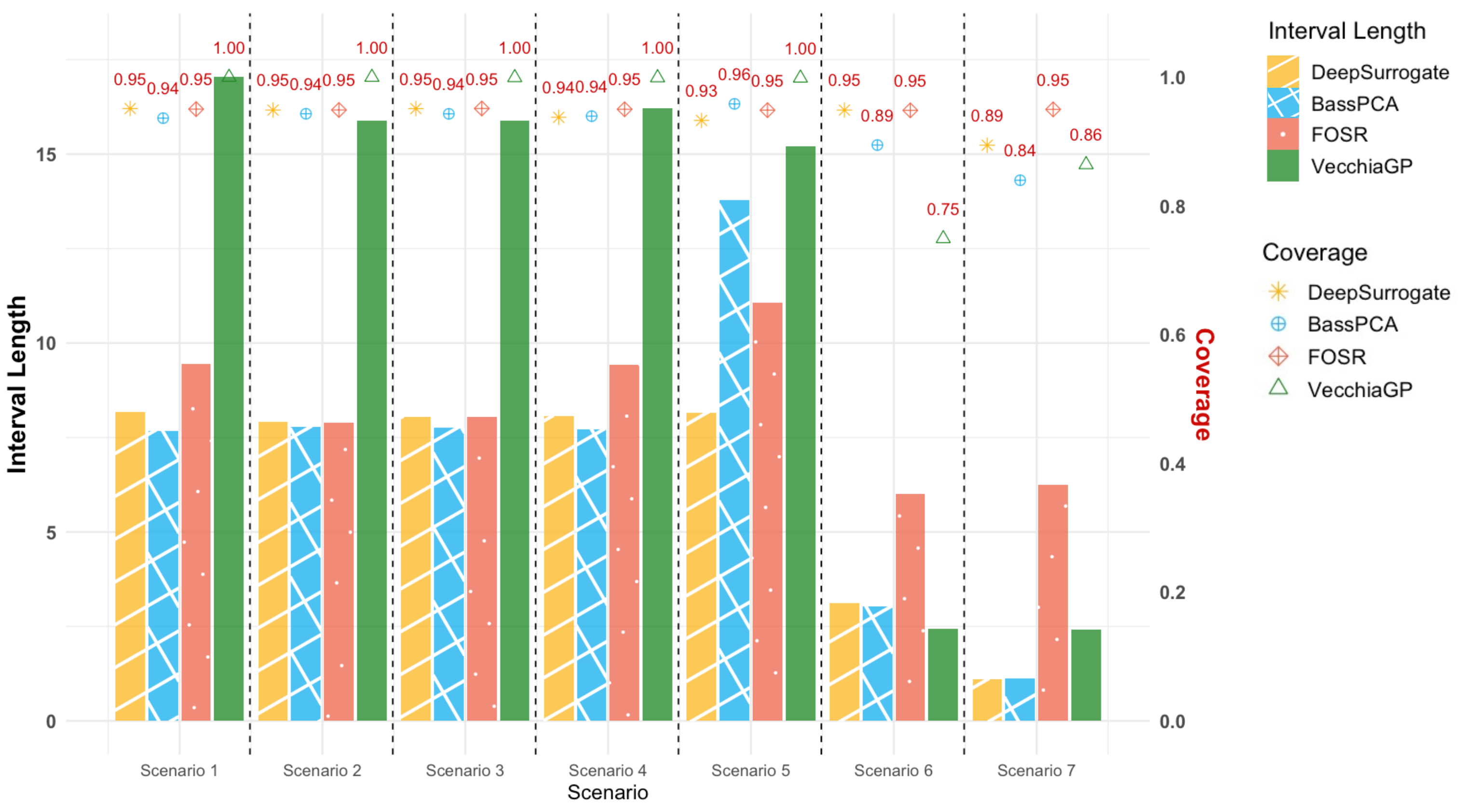}
    \caption{Comparison of predictive inference across competing models (DeepSurrogate, BassPCA, VecchiaGP and FOSR) in Scenario 1-7. The figure presents the interval length using a bar plot with their corresponding coverage probabilities overlaid for each scenario. DeepSurrogate shows near nominal-coverage in all scenarios, while other competitors either overestimates or underestimates predictive uncertainty.}
\label{fig:simul_y}
\end{figure}

According to Figure~\ref{fig:simul_y}, DeepSurrogate consistently achieves approximately nominal coverage across all scenarios, providing well-calibrated predictive intervals. As the number of simulations decreases, the stability of BassPCA deteriorates, leading to greater variability. Specifically, in Scenario 5, where the number of simulations is limited ($H=10$) and the SNR is low, BassPCA generates predictive intervals (PIs) that are nearly twice as wide as those produced by DeepSurrogate. As SNR increases in Scenarios 6 and 7, BassPCA exhibits slight under-coverage, with predictive intervals narrower than those of DeepSurrogate. Under low SNR conditions (Scenarios 1--5), VecchiaGP tends to over-cover, producing PIs roughly twice as wide as DeepSurrogate's, but shifts toward under-coverage in Scenarios 6 and 7, as SNR improves. FOSR performs similarly to DeepSurrogate in Scenarios 1--3 with abundant simulations; however, as the number of simulations declines, FOSR maintains near-nominal coverage at the cost of substantially wider intervals. Overall, these results highlight DeepSurrogate's competitive performance, particularly under higher SNR and the practical constraint of limited simulations common to high-fidelity computer experiments.

\subsection{Empirical Studies Under Mis-specified Data Generation Model}
The empirical results in Sections 4.1--4.3 were based on the assumption that the true data-generating process follows the basis expansion model specified in (\ref{basis_expansion}) to induce non-linear effects of the input vectors. This section assesses the performance of DeepSurrogate and its competitors under model mis-specification. Specifically, we simulate input attributes and fine-scale covariates as described in Section 4.1, but generate the function $f_{\bs}(\bz_h)$ from a Gaussian process with zero mean and an exponential covariance kernel. This kernel is defined by a spatial variance parameter $\alpha^{*2}$ and spatial scale parameter $l^*$, such that $Cov(f_{\bs_i}(\bz_h),f_{\bs_j}(\bz_{h'}))=\alpha^{*2}\exp(-\sqrt{||\bs_i-\bs_j||^2+||\bz_h-\bz_{h'}||^2}/l^*)$.  For each $h=1,..,H+H_0$ and $i=1,..,n$, the response $y_h(\bs_i)$ is drawn independently from $N(\beta_0^*+\bx(\bs_i)^T\bbeta^*+f_{\bs_i}(\bz_h),\delta^{*2})$ following the additive structure in (\ref{additive_local_global}). The true values of $\beta_0^*,\bbeta^*$ are set as in Section 4.1.

We consider four simulation scenarios described below. Given that the spatial variance and correlation have less impact on the relative performances of the competitors, we focus on two cases by varying the number of simulations and the noise variance $\delta^{*2}$ which mostly affect the relative performances. Since the main focus of this article is developing surrogates for high-fidelity simulations, we restrict ourselves to small values of $H$.
\begin{itemize}
    \item \textbf{Scenario 1}: $n=1000$, $H=15$, $H_0=5$, $\alpha^{*2}$ are drawn uniformly between $[1, 5]$, and $l^*$ is drawn uniformly between $[1, 5]$, $\delta^{*2}=1$.
    \item \textbf{Scenario 2}: $n=2000$, $H=6$, $H_0=4$, $\alpha^{*2}$ are drawn uniformly between $[1, 5]$, and $l^*$ is drawn uniformly between $[1, 5]$, $\delta^{*2}=1$.
    \item \textbf{Scenario 3}: $n=1000$, $H=15$, $H_0=5$, $\alpha^{*2}$ are drawn uniformly between $[1, 5]$, and $l^*$ is drawn uniformly between $[1, 5]$, and $\delta^{*2}=0.5$
    \item \textbf{Scenario 4}: $n=2000$, $H=6$, $H_0=4$, $\alpha^{*2}$ are drawn uniformly between $[1, 5]$, and $l^*$ is drawn uniformly between $[1, 5]$, and $\delta^{*2}=0.5$

\end{itemize}

\begin{table} [h!]
\small
\setlength{\tabcolsep}{4pt}
\centering
\begin{tabular}{l c c c c}
& \textbf{DeepSurrogate} &\textbf{BassPCA} &\textbf{FOSR} &\textbf{VecchiaGP}  \\
& &  &  & \\
\hline
\hline
\textbf{Scenario 1} & 2.3242 & \textbf{2.1746} & 3.0531 &  2.3430 \\
\textbf{Scenario 2} & \textbf{2.3186} & 4.5696  &  2.8357 & 2.3623  \\
\textbf{Scenario 3} & 1.0680 & \textbf{0.9530}  & 2.2234  & 0.9558  \\

\textbf{Scenario 4} & 0.6777 & 0.7468  & 2.0478  & \textbf{0.6651} \\

\hline
\hline
\end{tabular}
\caption[]{Predictive point estimates for DeepSurrogate and competing methods (BassPCA, FOSR, and VecchiaGP) across the four simulation scenarios under model mis-specification are presented. For each method, root mean squared prediction error (RMSPE) is reported. The best performing model under each scenario is boldfaced.}\label{table_simul_GP}
\end{table}
The results from Scenarios 1 and 2 under model misspecification largely mirror the conclusions drawn from the earlier simulations assuming correct model specification (refer to Table~\ref{table_simul_GP}). In Scenario 1, with a sample size $n=1000$ and number of simulations set to $H=15$, DeepSurrogate and VecchiaGP achieve comparable point prediction accuracy, while BassPCA shows a slight advantage. In contrast, when the sample size increases to $n=2000$ and the number of simulations is reduced to $H=6$, the primary setting of interest in this study, DeepSurrogate delivers the most accurate point predictions, with VecchiaGP performing competitively and FOSR showing notably poorer performance. Increasing SNR in Scenarios 3 and 4 show competitive point prediction of DeepSurrogate with VecchiaGP, BassPCA being close.

\begin{figure}[h!]
\centering\includegraphics[width=\linewidth]{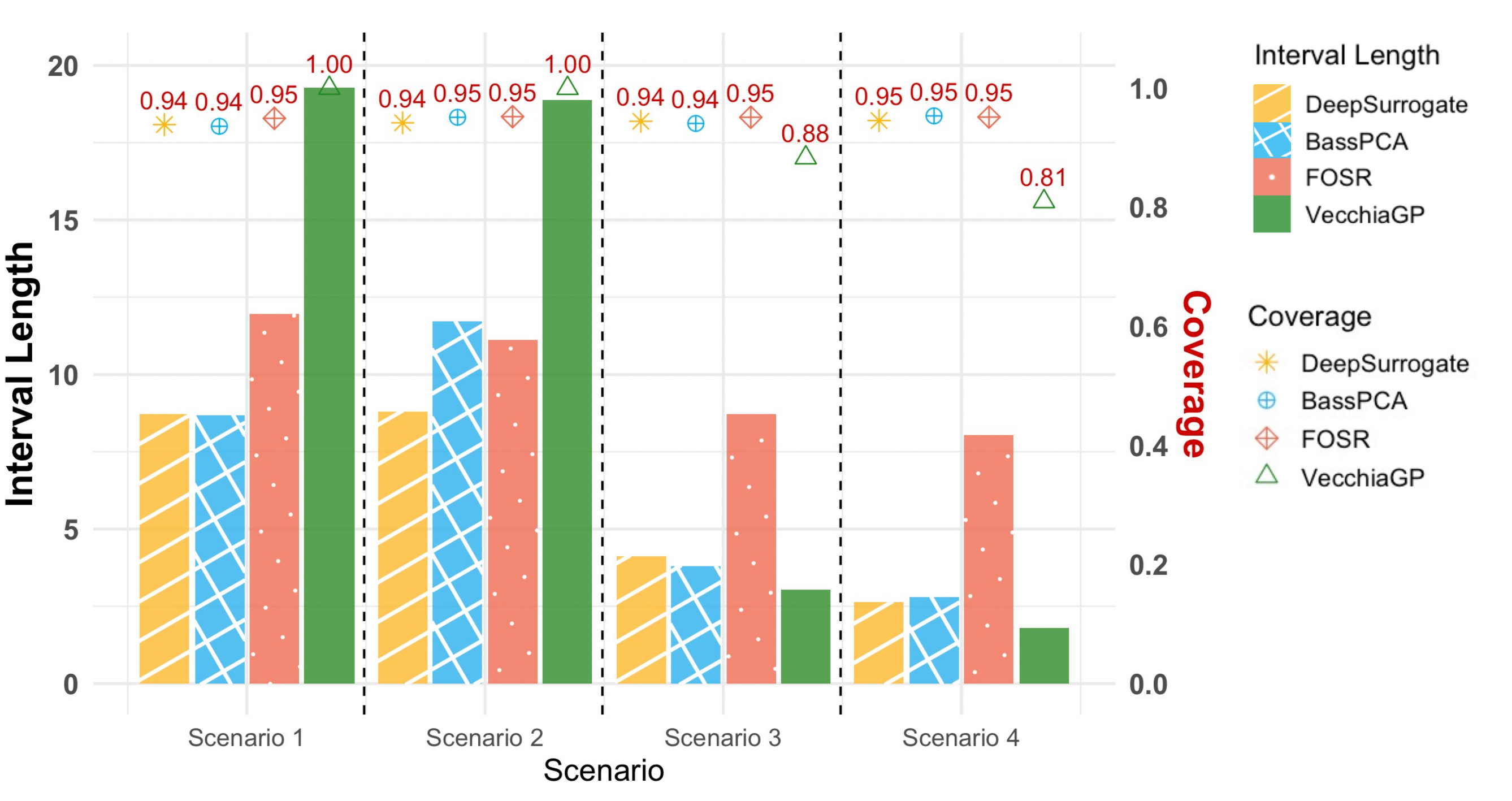}
    \caption{Comparison of predictive inference across competing models (DeepSurrogate, BassPCA, VecchiaGP and FOSR) in Scenario 1-4 under model mis-specification. The figure presents the interval length using a bar plot with their corresponding coverage probabilities overlaid for each scenario. DeepSurrogate shows near nominal-coverage in all scenarios, while other competitors either overestimates or underestimates predictive uncertainty.}
\label{fig:simul_y_GP}
\end{figure}

As illustrated in Figure~\ref{fig:simul_y_GP}, DeepSurrogate consistently delivers well-calibrated predictive intervals with coverage probabilities close to the nominal level across all scenarios. FOSR also achieves near-nominal coverage but produces substantially wider predictive intervals than DeepSurrogate. BassPCA yields comparable uncertainty quantification in Scenarios 1 and 3 when the number of simulations is $H=15$, but exhibits much wider predictive intervals in Scenarios 2 and 4. Under low SNR conditions in Scenarios 1 and 2, VecchiaGP displays over-coverage, with predictive intervals nearly twice as wide as those from DeepSurrogate. As the SNR increases, however, VecchiaGP tends to under-cover, providing narrower intervals---a pattern also noted in Section 4.3.

\section{SLOSH Emulator Data Analysis}\label{slosh}
This section presents the application of our Artificially Intelligent Functional Surrogate Model, referred to as DeepSurrogate, to analyze storm surge simulation data alongside competing models. Storm surge models play a crucial role in emergency response, planning, and research by simulating floodwater depths caused by hurricanes. Among these, the Sea, Lake, and Overland Surges from Hurricanes (SLOSH) simulator \citep{jelesnianski1992slosh}, developed by the National Weather Service, is one of the most widely used models. See \cite{hutchings2023comparing} for previous work in SLOSH emulation.

Since the goal is to test efficacy of DeepSurrogate under limited simulations for its potential usage in expensive simulation settings, we analyze a dataset consisting of $30$ simulations generated using the SLOSH simulator, each representing a distinct hypothetical storm scenario. These storms are defined by unique combinations of five input parameters. Four parameters describe the hurricane at landfall: the heading of the eye, the velocity of the eye, the latitude of the eye, and the minimum air pressure. The fifth parameter, projected sea level rise for the year 2100, introduces an element of uncertainty into the system.

\begin{figure}[h]
    \centering
    \includegraphics[width=0.85\linewidth]{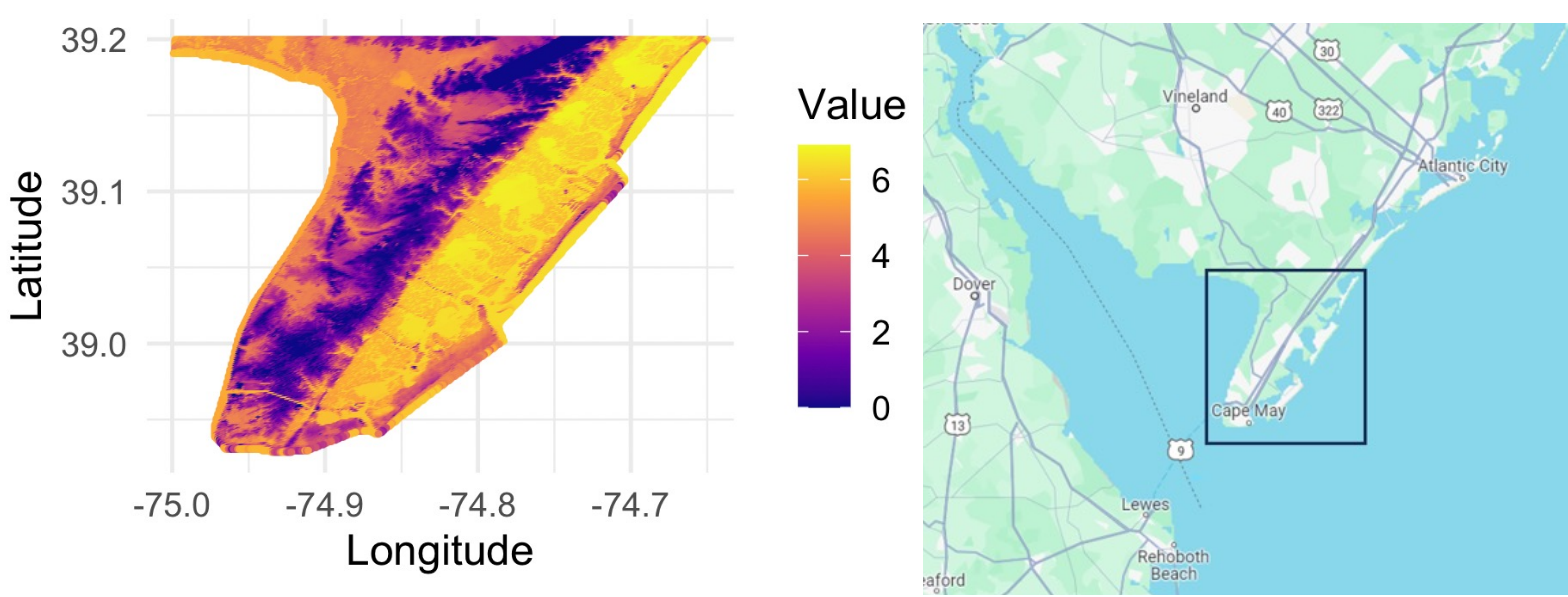}
    \caption{Output from one SLOSH model run, side-by-side with the location of the peninsula in Google Maps. Here, $x$ and $y$ axes represent longitude and latitude, respectively.}
\label{fig:locations}
\end{figure}

The parameter ranges, outlined in Table \ref{tab:ranges}, reflect realistic variations and were used to generate the simulation scenarios. This ensemble serves as a testbed for developing functional surrogates that enable efficient analysis and prediction in the context of computationally expensive storm surge simulations. The primary focus of this study is assessing storm surge-induced damage to electrical power stations. Since most power stations are located inland and remain unaffected, we specifically analyze functional data consisting of simulated storm surge heights at
$n=49,719$ spatial locations to predict hurricane-induced flooding in the southern tip of New Jersey (see Figure~\ref{fig:locations}) for each simulation. A map illustrating the output from a single SLOSH simulation over the spatial domain of interest is provided in Figure \ref{fig:locations}. Additionally, elevation data is available for each spatial location of the electrical power stations. Given the critical role of this covariate in assessing maximum flood heights at specific locations, we incorporate elevation as a \emph{fine-scale covariate} in our analysis to enhance predictive accuracy.

\begin{table}
\small
\setlength{\tabcolsep}{4pt}
    \centering
    \begin{tabular}{lccccc}
    \toprule[1.5pt]
        Predictor & Heading & Velocity & Latitude & Pressure & Sea level rise (2100) \\
        \midrule
       Lower & 204.0349 & 0 & 38.32527 & 930 & -20\\
       Upper & 384.0244 & 40 & 39.26811 & 980 & 350\\
       Units & degrees & knots & degrees & millibars & cm\\
       \bottomrule[1.5pt]
    \end{tabular}
    \caption{Parameters varied in SLOSH simulations.}
    \label{tab:ranges}
\end{table}

To evaluate the out-of-sample predictive performance of the competitors, we utilize 10 simulation datasets for model fitting, $10$ simulations for validations and $10$ simulations from the SLOSH simulator for testing. Given that each simulation contains data for $49,719$ locations, this yields predictive inference on $n_{\text{test}}=497,190$ out-of-sample observations.

\noindent \underline{\textbf{Mis-classification error for power stations}}. Power stations in the Delaware Bay region are typically designed to withstand flooding up to four feet, beyond which severe damage becomes inevitable. Therefore, we also evaluate the emulator's ability to accurately predict whether a storm surge exceeds this critical threshold. Specifically, we report the percentage of misclassification, where a misclassification occurs when a true storm surge height above four feet is incorrectly predicted as below four feet, or vice versa. This predictive capability is crucial for informed decision-making, enabling timely interventions such as pre-emptively shutting down power stations in preparation for an approaching storm.

\subsection{Results}
\begin{figure}[h]
\begin{center}
\includegraphics[width = 0.9\textwidth]{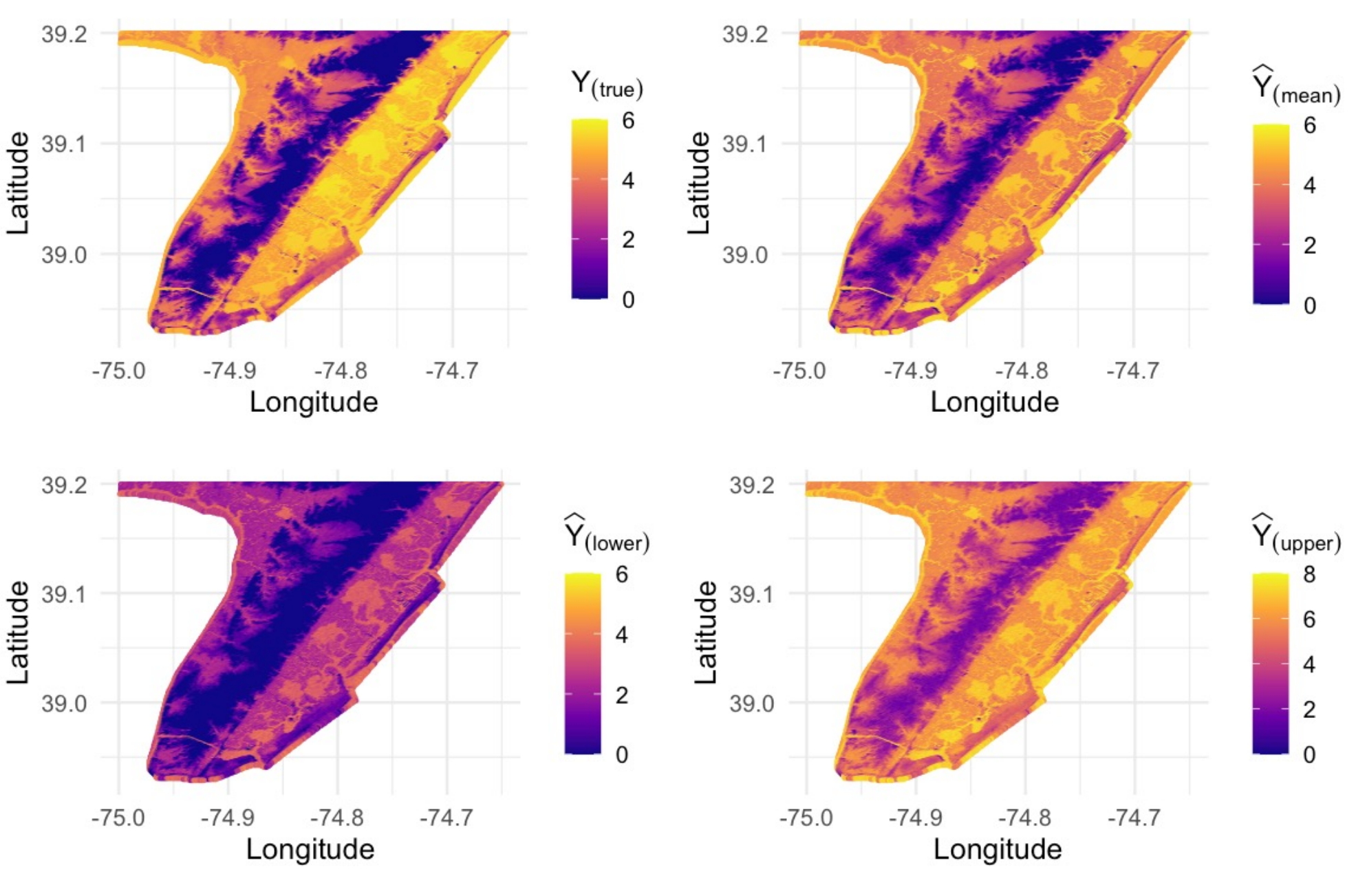}
\end{center}
\caption[]{The left-upper figure and the right-upper figure show the true simulated storm surge and the estimated storm surge in a randomly selected out-of-sample simulation. The left-bottom and the right-bottom figures provide 2.5\% and 97.5\% predictive interval limits.}
\label{fig:results_simul10}
\end{figure}

Figure~\ref{fig:results_simul10} displays the true and predicted storm surge patterns for a randomly selected test storm, illustrating that DeepSurrogate effectively captures fine-scale spatial variability in storm surge. As summarized in Table~\ref{tab:storm_surge}, both DeepSurrogate and VecchiaGP demonstrate strong point prediction performance, with comparable RMSPE and mis-classification rates. However, in terms of uncertainty quantification, DeepSurrogate shows a clear advantage by producing well-calibrated 95\% predictive intervals with coverage close to the nominal level, whereas VecchiaGP tends to under-cover due to overly narrow intervals. Additional analysis estimates the signal-to-noise ratio in the storm surge data to be 3.86, which aligns closely with Scenario 7 from Section 4.1 and Scenarios 3 and 4 from Section 4.4. Consistent with the results from these scenarios, DeepSurrogate strikes a more effective balance between predictive accuracy and uncertainty quantification than VecchiaGP.

Aligned with the simulation findings, BassPCA struggles under limited simulation settings, exhibiting higher RMSPE and misclassification rates relative to both DeepSurrogate and VecchiaGP. FOSR performs the worst in terms of point prediction accuracy, while both FOSR and BassPCA generate overly wide predictive intervals with 100\% coverage, reflecting excessively conservative uncertainty estimates.

\begin{table}[ht!]
\small
\setlength{\tabcolsep}{4pt}
    \centering
    \begin{tabular}{lccccc}
        \toprule
        & \textbf{RMSPE} &  \textbf{Error \%} & \textbf{Coverage} & \textbf{Length} & \textbf{Comp. time (in min.)}\\
        \midrule
        $\text{BassPCA}$ & 1.81 & 0.18 & 1.00 & 11.51 & 0.41 \\
        $\text{VecchiaGP}$ & \textbf{0.78} & \textbf{0.11} & 0.80  & 2.01  & 2.36 \\
        $\text{FOSR}$ & 2.14 & 0.47 & 1.00 & 8.28 & 0.37 \\
       $\text{DeepSurrogate}$ & 0.90 & 0.12 & 0.94 & 3.44 & 8.03\\
        \bottomrule
    \end{tabular}
    \caption{The table summarizes the predictive performance of DeepSurrogate and its competitors---VecchiaGP, BassPCA, and FOSR---across 10 out-of-sample simulations. It reports the root mean squared prediction error (RMSPE), misclassification error rate for power stations (Error\%), coverage of 95\% predictive intervals, average length of 95\% predictive intervals, and computation time (in minutes) for all methods.}
    \label{tab:storm_surge}
\end{table}

\section{Conclusion and Future Work}
This article presents DeepSurrogate, an explainable AI framework designed for modeling functional surrogates with vector-valued inputs in physics-based, high-fidelity simulations. The framework effectively captures spatial dependencies in the functional output, models complex relationships between inputs and outputs, and incorporates MC-dropout for principled uncertainty quantification---all within a deep neural network (DNN) architecture. This integration enhances the interpretability of DNNs and contributes to the expanding landscape of interpretable machine learning methods. In settings where simulations are costly and limited in number, DeepSurrogate offers accurate predictions with well-calibrated uncertainty estimates. It is competitive in point prediction with popular alternatives while offering superior uncertainty quantification. Moreover, it remains computationally efficient, enabling full predictive inference for 50,000 spatial locations in under 10 minutes.

While our framework has been employed to develop efficient surrogates, it has important applications in statistical imaging. As a natural extension of our work, we aim to develop an AI-based generative modeling framework for image-on-image regression where input and output images are structurally misaligned. In this setting, spatial dependencies will play a crucial role; however, additional topological structures, such as network-based associations between image regions, may also influence the modeling process.
In an ongoing work, we are exploring methodologies that integrate both spatial and network dependencies in images.

\section{Acknowledgment}
Dr. Rajarshi Guhaniyogi is supported by National Science Foundation DMS-2210672, and National Institute of Health R01NS131604.

\section*{Disclosure Statement}
No potential conflict of interest was reported by the authors.

\section*{Data Availability Statement}

Deidentified data have been made available at the following URL:\url{https://dataverse.harvard.edu/dataset.xhtml?persistentId=doi:10.7910/DVN/7K5O5X}.

\bibliographystyle{abbrvnat}
\bibliography{references_new}
\end{document}